# Polymer Sequence Design via Active Learning

Praneeth S Ramesh and Tarak K Patra[*]

Department of Chemical Engineering, Center for Atomistic Modeling and Materials Design and Center for Carbon Capture Utilization and Storage, Indian Institute of Technology Madras, Chennai, TN 600036, India

**Abstract:**

Analysis of molecular scale interactions and chemical structure offers an enormous opportunity to tune material properties for targeted applications. However, designing materials from molecular scale is a grand challenge owing to the practical limitations in exploring astronomically large design spaces using traditional experimental or computational methods. Advancements in data sciences and artificial intelligence have produced a host of tools and techniques that can facilitate the efficient exploration of large search spaces. In this work, a blended approach integrating physics-based methods and data science techniques is implemented in order to effectively screen the materials configuration space for accelerating materials design. Here, we survey and assess the efficacy of data-driven methods within the framework of active learning for a challenging design problem viz, sequence optimization of copolymers. We report the impact of surrogate models, kernels, and initial conditions on the convergence of an active learning method for the sequence design problem. This work outlines optimal strategies and parameters for inverse design of polymer sequence via active learning.

*Corresponding author, E-mail: tpatra@iitm.ac.in

# 1. Introduction

Copolymers are a class of macromolecules that are prepared by combining two or more different types of monomers using appropriate polymerization and synthesis techniques. Optimal properties of copolymers intended for specific applications, such as interfacial energy, selective chemical transport behaviour (proton, ion or water transport), structural phase transitions, kinetics of bio-degradation, drug release profiles, have been found to be strongly dependent on the precise sequence of the copolymer chain.[1,2] In many cases, this sequence-specificity is so powerful that a subtle change in the copolymer sequence results in a huge change in the property of interest.[3] At times the optimal property that one is seeking for is present in a non-intuitive, seemingly arbitrary polymer sequence, the sequence-specificity of which cannot be approximated by coarse sequence statistics or mean-field theories. The observation that sequence-specificity is crucial for property optimization is well known in biological systems, wherein sequence-specific macromolecules such as proteins, lipid bilayers, ribonucleic acids and deoxy-ribonucleic acids have been optimized by nature for their functionalities over millions of years of natural evolution[4]. Even as sequence-specific property control offers great scope for obtaining tunable properties by designing sequences that are novel and hitherto-unheard of, a systematic approach for identifying such "optimal sequences" is an enormous task. Intuition-based material design approaches / rudimentary trial-and-error approaches would fetch limited results[2], while on the other hand brute-force approaches enumerating the entire range of sequence possibilities and measuring/computing their properties are time-consuming, laborious, inefficient and impractical. Efforts involved in property measurement through experimentation or high-throughput computation can be quite resource-intensive, be it manpower, experimental costs or computational processing power. Even for a simple linear copolymer chain of AB type of chain length 100, with 2 types of chemical moieties, brute-force enumeration would result in $2^{99}$ sequences which is over $10^{29}$ possibilities. The sequence possibilities would be exceedingly higher for longer copolymer systems comprising several monomer constituents. A feasible and viable approach is therefore necessary for intelligently identifying sequences that are more likely to have a particular desired property; and the available resources, time and effort could be channelized towards a smaller set of sequence configurations, rather than exhaustively scanning the entire sample space. Such an efficient exploration of vast search spaces[5] can be done with the help of certain tools and techniques that have emerged recently, in the wake of latest advancements in data science and artificial intelligence. A blended approach that integrates physics-based methods and data science, by incorporating both first-principles calculations using traditional Molecular Mechanics approaches and Machine Learning based techniques, offers an efficient way for screening the materials configuration space and thereby accelerating materials design.

This work explores a representative case study on 100-mer AB type binary copolymer sequences. The objective is to identify the precise copolymer sequence with the desired target property of 'lowest Radius of Gyration' in a solvent environment. Coarse-grained molecular dynamics simulations are used for the purpose of estimation of radius of gyration, taking into account the slow process of polymer relaxation, and the time-consuming nature of characterizing the equilibrium properties of polymers. While the use of high-throughput molecular dynamics computations instead of experiments is much faster; nevertheless, it is necessary that we scan the sequence space far more efficiently with minimal number of computationally expensive MD simulations. Since the desired objective for the target property is well-defined but the corresponding optimal sequence is unknown, this problem falls under the purview of inverse design problems. In this regard, approaches such as Evolutionary Algorithms, Monte Carlo Tree Search, Generative models using unsupervised learning techniques have been reported as standard tools for tackling inverse design problems.[5] However, the concept of utilizing conditional probability for scanning the extensive combinatorial search space makes Active Learning using Bayesian Optimization an enviable strategy, in comparison to other techniques. While all these different techniques involve sampling newer data points in an iterative manner, the rationale[6] behind the strategic selection of the newer data points is best captured by the Bayesian optimization strategy.

Active learning strategy within the Bayesian Optimization framework is implemented to streamline the cherry-picking of candidate sequences for running through expensive computations and this accelerates the search for optimal copolymer sequence candidates that correspond to the global maxima or global minima of Radius of Gyration. Fingerprinting of the 100-mer AB type copolymer sequence is done by coding the As by 1s and Bs by 2s, and this defines the input variable, while the target property is defined by the radius of gyration. This first step of the active learning framework involves estimating the objective function through a surrogate model implemented over the training set. The premise is to obtain a proxy for the objective function through a computationally cheap pathway. Machine-Learning (ML) models such as Gaussian Process Regression (GPR), Support Vector Regression (SVR) and Kernel Ridge Regression (KRR) lend themselves conveniently for this purpose.[7] These surrogate models can now be used to make computationally cheap predictions for Radius of Gyration on other copolymer sequences in the search space of sequences for which Radius of Gyration values have not been computed using Molecular Dynamics. In this regard, it is important to note that the surrogate predictive model must necessarily provide uncertainty measures for every prediction, as this is the hallmark of the active learning approach.[8] The uncertainty, in some sense, is a quantitative description of the degree to which a new sequence falls within/outside the domain of the training dataset. The estimates of the prediction values and the uncertainty values are crucial in selecting the candidate copolymer sequences for which computation of Radius of Gyration would be the most strategic. This is the second step of the Active learning, in which a select few cherry-picked sequence candidates are recommended for computation through expensive molecular dynamics simulations. The cherry-picked polymer sequences and the associated Radius of Gyration values are appended to the training set, and the surrogate model is retrained using a slightly larger training set. The aforementioned steps are repeated iteratively, and a progressively expanding training set is generated so as continuously update and refine the surrogate model. ML models usually operate along the framework, "greater the data, better is the model prediction", and therefore it is expected that the search for optimal values becomes easier in subsequent iterations, even if the model quality is poor to start with[9].

The different nuances of the Bayesian Optimization approach, such as the choice of surrogate model and associated kernel, selection of hyper-parameters in surrogate models, number of iterations required for convergence, stopping criterion, rate of progress towards convergence, model quality across iterations are analysed for a problem whose solution is intuitively known. Following this, an AB-type copolymer system of chain length 100 with the number of As and Bs constrained to be in a 50:50 ratio is considered, and the active learning framework is implemented to find the optimal sequence(s) with the least radius of gyration value. As it is impossible to intuitively identify the optimal sequence configuration for this problem, the capability of the Active Learning framework with Bayesian Optimization to identify such non-intuitive optimal sequences is successfully demonstrated.

## 2. Methodology

### 2.1 Model Polymer

We consider a generic coarse-grained model of a copolymer to explore its sequence. In this model system, two adjacent coarse-grained monomers of a polymer are connected by the Finitely Extensible Nonlinear Elastic (FENE) potential of the form $E = -\frac{1}{2}KR_0^2 \ln\left[1 - \left(\frac{r}{R_0}\right)^2\right]$, where $K = 30\epsilon/\sigma^2$ and $R_0 = 1.5\sigma$. The pair interaction between any two monomers is modelled by the Lennard-Jones (LJ) potential of the form $V(r) = 4\varepsilon\left[\left(\frac{\sigma}{r}\right)^{12} - \left(\frac{\sigma}{r}\right)^6\right]$. The $\epsilon$ is the unit of pair interaction energy. The size of all the monomers is σ. The LJ interaction is truncated and shifted to zero at a cut-off distance $r_c = 2.5\sigma$ to represent attractive interaction among the monomers. The AA interaction strength is $\epsilon_{AA} = \epsilon$, while the BB and AB type interaction strengths are $\epsilon_{BB} = \epsilon_{AB} = 0.2\epsilon$. This ensures immiscibility between A and B type monomers. These set of model parameters are previous used to study copolymer systems, and are very successful in understanding the generic properties of both synthetic polymers and biomolecules. We choose a polymer of chain length N=100.

### 2.2 Molecular Dynamics Simulations

We conduct implicit solvent molecular dynamics simulations of a polymer chain in a canonical ensemble. The initial configuration of a polymer chain is placed in a cubic simulation box of fixed dimension and box is periodic in all three directions. We use the Velocity Verlet algorithm with a timestep of $0.001\tau$ to integrate the equation of motion. Here, $\tau = \sigma\sqrt{m/\epsilon}$ is the unit of time, and $m$ is the mass of a monomer, which is same for both the A and B type moieties. All the simulations are conducted at a reduced temperature $T^* = T\,\epsilon/k_B = 1$, which is maintained by the Langevin thermostat within the LAMMPS simulation environment. The simulations are conducted for polymer chains of length N= 100. All the simulations are equilibrated for $10^7$ MD steps followed by a production run of $10^7$ steps. The data during the production cycles are collected for computing the radius of gyration of a polymer chain.

### 2.3 MD based Active Learning

Bayes' theorem describes the posterior distribution $P(f|D)$ of an event based on its prior probability and likelihood function, which can be written as

$$P(f|D) = \frac{P(D|f)P(f)}{P(D)}$$

where $P(f)$ stands for the prior probability distribution and $P(D|f)$ stands for the likelihood function. The denominator $P(D)$ is the marginal likelihood or evidence, which is usually computationally intractable, and is merely a normalizing constant. Here, D is the given data set. In essence, $P(f|D) \propto P(D|f)P(f)$. The prior *P(f)* represents the *a priori* belief about the space of possible objective functions before any data is taken into consideration. Although the precise functional detail about f remains unknown without any data, prior knowledge about some of the properties of the objective function such as the assumption of its smoothness are captured by the prior function. As we accumulate data D {X, Y}, where X refers to the set of copolymer sequences and Y refers to the corresponding Radius of Gyration, the prior distribution is combined with the likelihood function *P(D|f)* to obtain the posterior which captures our updated beliefs about the unknown objective function (the Radius of Gyration).[10]

Materials design framework can be established based on the Bayesian approach of relating posterior distribution and prior probability. As shown in Figure 1, a training set $X_{train}$ is defined by randomly selecting 100 copolymer sequences from the vast combinatorial space of $2^{99}$ sequences. For each of the copolymer sequence that is part of $X_{train}$, the corresponding radius of gyration value is determined through MD simulations so as to populate the corresponding $Y_{train}$. By building a surrogate model over the initial training set of $\{X_{train}, Y_{train}\}$, computationally cheap predictions of Radius of Gyration can be obtained for other copolymer sequences whose Radius of Gyration values are unknown. The estimation of the prediction values by a surrogate model and the quantification of the degree of uncertainty associated with each prediction are crucial elements of the Active Learning framework. We have used Gaussian process Regression (GPR), Support Vector Machine (SVM) and Kernel Ridge Regression (KRR) as as surrogate models in this work. The GPR provide an estimation of the uncertainty of a prediction.[11] However, SVM and KRR do not estimate uncertainties, and thus a bootstrap approach[12] is employed for obtaining uncertainty measures. For each of these three surrogate models, we choose two kernels, resulting in a sum total of six models. The model architecture in each of these six models is dependent on a set of hyper-parameters that must be specified in advance, before training on a dataset. The choice of hyper-parameters plays a vital role in governing the quality of the surrogate model and in turn the performance of the optimizer.[13,14] Based on our initial study we choose a set of hyper-parameters which remains fixed during the active learning process. We report the set of hyperparameters for all the models in SI.

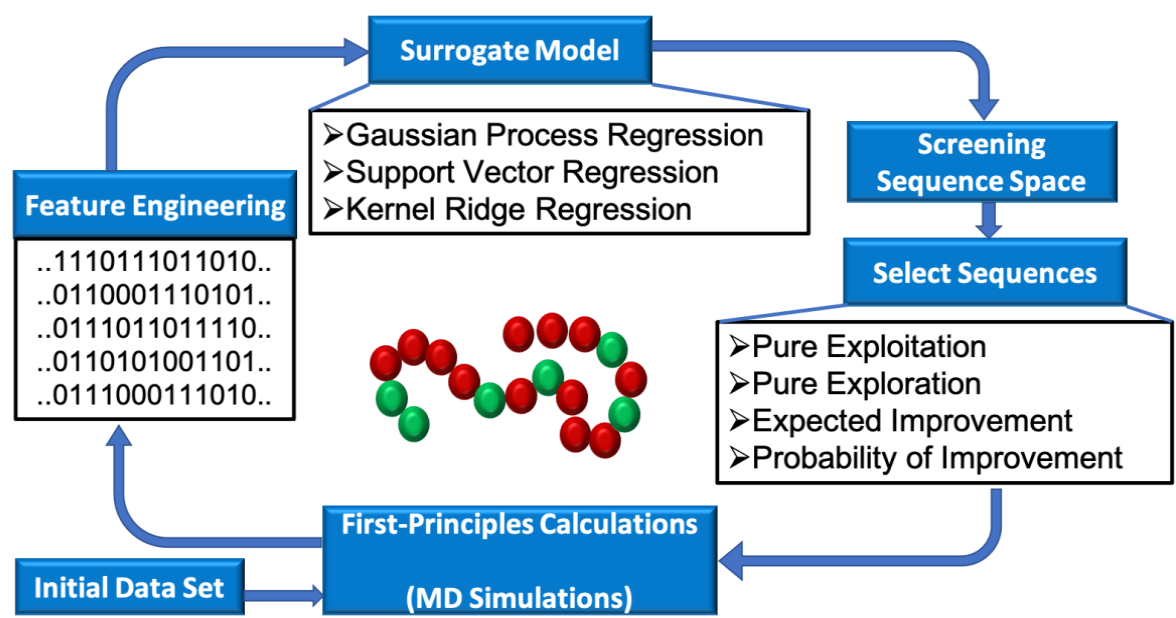

*Figure 1: Active Learning workflow. The initial set of sequences are randomly generated and their properties are calculated using MD simulations. The chemical moieties in these sequences of are mapped to a string of 1s and 0s for facilitating easier feature representation. Surrogate models are built based on these data. Subsequently, the sequence space is screened to select a predefined number of candidate sequences, based on the predictions of the surrogate models. MD simulations are then conducted for this set of selected candidates. The process continues until a termination criteria is reached.*

Now, once the hyper-parameters of a surrogate model are chosen and subsequently, when the surrogate model has been trained over a training set, it is computationally expensive to run this surrogate model to make predictions over the entire space of copolymer sequences whose Radius of Gyration values are unknown. Instead, a candidate search space is defined by considering the copolymer sequences that are sequentially proximate to the training set (more details of this are discussed in the subsequent section). Predictions and their uncertainties are obtained for all the sequences in the selected candidate search space. A subset of these candidates are shortlisted for first principles-based property estimation. Here we choose 32 candidates for computing radius of gyration via MD simulation. This selection is done according to different query strategies namely pure exploitation, pure exploration, and using acquisition functions such as Expected Improvement and Probability of Improvement. A random selection of sequences is also performed to serve as a baseline to which the performance of the different query strategies could be compared with. With six models, and five query strategies (including the random selection), a total of 30 optimization frameworks are tested here for the sequence design problem. The Active Learning process comprising of different steps such as training the surrogate model on a training set, definition of a candidate search space, running the model to estimate predictions and uncertainties, selection of candidates based on a query strategy, computation of Radius of Gyration of the selected candidates by MD simulations, and the re-training of the surrogate model with the newly sampled sequences, is repeated in an iterative manner for a total of 75 iterations as shown in Figure 1. Thus, in every iteration, 32 new sequences and their associated Radius of Gyration values are appended to the training set. The mathematical details of the three surrogate models and their kernels are presented in the supplementary information (SI). In addition, the SI describes the five different query strategies used in this work.

## 3. Results and Discussion

First, we validate the model and the method that are used for calculating the radius of gyration of single chain copolymers. Figure 2a shows the pair energy and Rg of a randomly selected copolymer during the equilibration run followed by production run. The MD snapshots of the initial and final configuration is shown in Figure 2b and c, respectively. It is clear that the equilibrium is achieved during $10^7$ steps. The equilibrium energy and Rg value are around $2.82.8\epsilon$ and $3.1\sigma$, respectively. As mentioned in the methods section, this equilibrium run is followed by a production run of another $10^7$ steps during which average Rg value is estimated. For this particular chain, the radius of gyration is $3.10 \pm 0.15\sigma$. The statistical fluctuation in Rg is ~2%. This is the typical range of error bars in all our data.

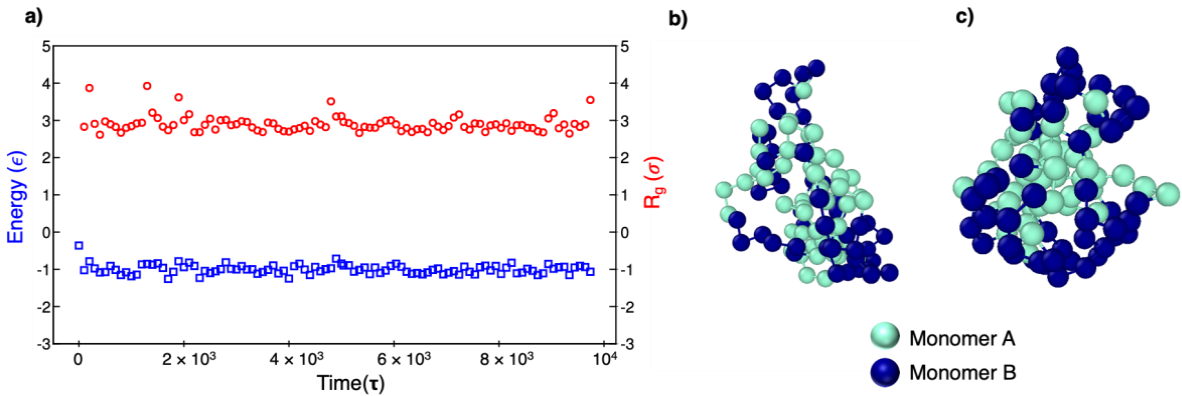

*Figure 2: Validation of model and method. (a) The radius of gyration and pair energy of the system during the equilibration run followed by production run. (b) The initial configuration of the polymer. (c) The final configuration of the polymer*

### 3.1 Performance of different Active Learning Strategies

Active learning of a copolymer's sequence-Rg correlation within the BO framework is carried out over 75 iterations using six different predictive models (three surrogates with two kernels each) and five query strategies in each model. In each iteration, a total of 32 candidates' Rg values are directly measured using MD simulations. The initial data set is same for all the case studies. Being a minimization problem, the performance of the optimizer is primarily assessed by monitoring the manner in which sequences with lower radii of gyration are identified with increasing MD computations.

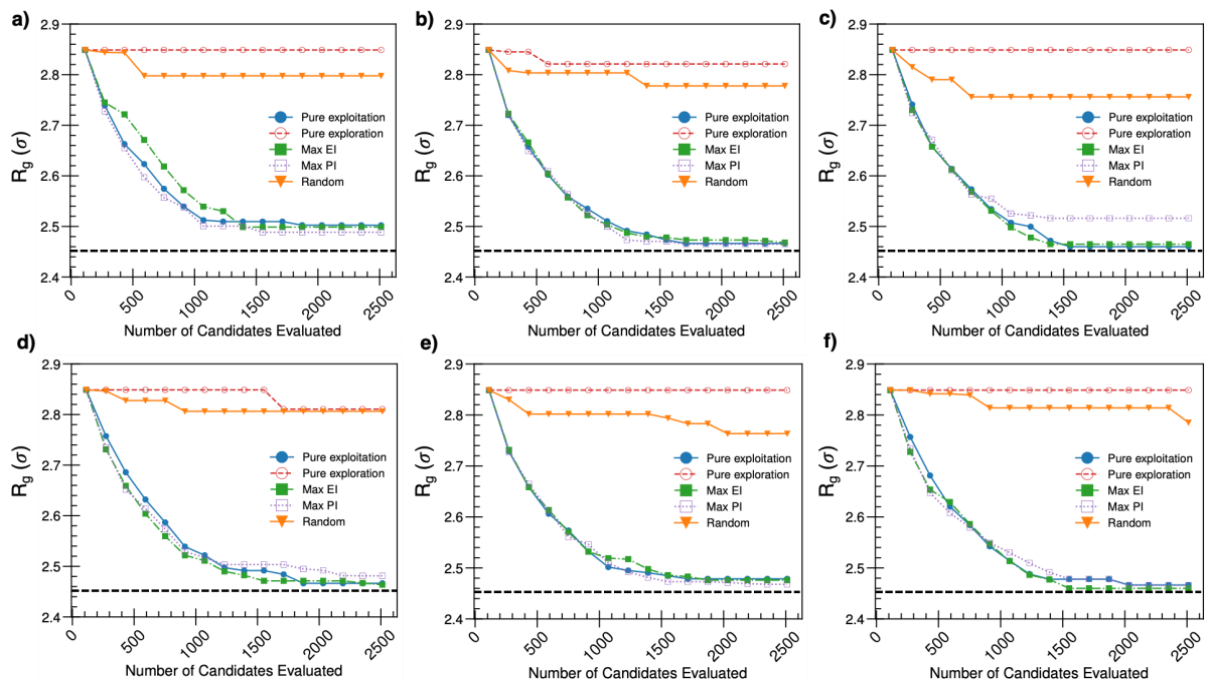

*Figure 3: Active learning of radius of gyration of the copolymer using six surrogate models. The performance of active learning via GPR-Rbf kernel, GPR-Matérn Kernel, SVR-Linear Kernel, SVR-Rbf kernel, KRR-rbf kernel, and KRR-Laplacian kernel is shown in (a), (b), (c), (d), (e) and (f), respectively. For each of these six strategies, five different queries are used as mentioned in all the subplots.*

Figure 3 shows cumulative plots which represent the minimum Rg observed so far, versus the number of MD computations performed so far for all the case studies. Superficially, two query strategies – pure exploitation, Max EI, in all cases (a to f) exhibit similar behaviour in approaching towards a Rg value around 2.48. Evidently, pure exploration and random queries in all the cases produces suboptimal solutions where the search is trapped in local minima for significant amount of time. The regions in the graph wherein the minimum radius of gyration is the same over a span of few iterations can be inferred to be regions/zones of local minima. The success of the Bayesian Optimization lies in its ability to circumvent local minima, by intelligently sampling newer sequences which facilitate the convergence towards global minima. Thus, it can be inferred that the Bayesian Optimization with the three query strategies – Pure Exploitation, Max EI and Max PI iteratively identifies candidates with a focus on minimizing the objective function, Radius of Gyration, the extent of decrease of which in each iteration can be observed to be far more pronounced compared to a Random Selection approach. On the other hand, with the pure exploration query strategy, in each of the six subcases, the Pure Exploration strategy shows poor convergence, with the Rg values being confined within a narrow range of 2.80 to 2.85. In a few subcases, it is even marginally outperformed by the Random Selection approach. These observations arise from the construct of this specific query strategy which focusses on expanding the sequence – Rg dataset without explicitly attempting to minimize the target property. For the purpose of critically assessing the performance of different query strategies, the number of iterations required to identify a sequence with an average Rg of 2.47 was considered as a quantitative measure. As far as GPR -Rbf kernel is concerned, with the max PI query strategy, the optimizer is stuck in a local optimum at 2.488. With the query strategies Max. EI and Pure Exploitation, the optimizer is stuck at a local optimum of 2.498 and 2.502 respectively. The results with GPR (Matérn Kernel) demonstrate the enhanced performance arising out of improved flexibility of a Matérn Kernel in comparison with an Rbf kernel. In the case of GPR (Matérn Kernel) for the query strategy max PI, after 1232 computations, the sequence corresponding to an Avg Rg value of 2.473 is identified, and subsequently by the end of 1584 computations, sequence corresponding to an Rg value of 2.466 is also identified. The query strategies pure exploitation and max EI result in a slightly delayed convergence to an Rg value of 2.473. In case of the SVR (Linear kernel), the max EI query strategy identifies a sequence with an Rg value of 2.473 after 1296 computations, while the pure exploitation does so after 1360 computations. The max PI query strategy results in a local minima trap at an Rg of 2.516. In case of SVR (Rbf kernel), max EI approach identifies a sequence with an Rg value of 2.471 after 1520 computations, while the pure exploitation identifies a sequence with an Rg of 2.473 after 1744 computations. The max PI query strategy in this case, does indeed result in a satisfactory optimization, but is trapped in a local optimum around 2.50 for quite a number of computations, before converging towards lower Rg values. In KRR (Rbf kernel), max PI query strategy identifies an Rg of 2.47 after 1488 computations while both max EI and pure exploitation result in a convergence towards an Rg of 2.47 after 1744 computations. KRR (Laplacian kernel) with pure exploitation query strategy identifies a sequence with Rg of 2.477 after 1264 computations while max EI and max PI identify a similar sequence after 1328 and 1456 computations respectively. In essence, the number of MD simulations for convergence towards an Rg value of 2.47 is summarized in Table 1.

| Surrogate Model / Query Strategy | GPR | | SVR | | KRR | |
|---|---|---|---|---|---|---|
| | Kernel | | Kernel | | Kernel | |
| | Rbf | Matérn | Linear | Rbf | Rbf | Laplacian |
| Pure Exploitation | - | 1552 | 1360 | 1744 | 1744 | 1264 |
| Max. EI | - | 1712 | 1296 | 1520 | 1744 | 1328 |
| Max. PI | - | **1232** | - | - | 1488 | 1457 |

*Table 1: Number of MD simulations to identify a sequence with an Rg of 2.47 for different surrogate models and query strategies. All the case studies begin with same set of initial data points.*

Now, we focus on the quality of solution by counting the number of A and B type moieties of the sequences that are learned iteratively. Figure 4 plots the number of As corresponding to the cumulative minimum Rg value (depicted in Figure 3) versus the number of computations performed so far. The number of computations after which sequences with number of As greater than or equal to 95 are consistently sampled correlates with the number of computations corresponding to convergence towards an Rg value of 2.47. Ideally the minimum Rg is present in a sequence with all the 100 monomeric units as As. At the end of 2512 computations, the Bayesian Optimizer has been able to identify sequences with 99 As, but not 100. This is not a serious handicap, as the Rg values for sequences with number of As ranging from 95 to 100 are expected to be roughly similar, owing to the thermal noise in the Molecular Dynamics implementation. Perhaps, the active learning framework might possibly narrow down on the sequence with 100 As after a few more iterations.

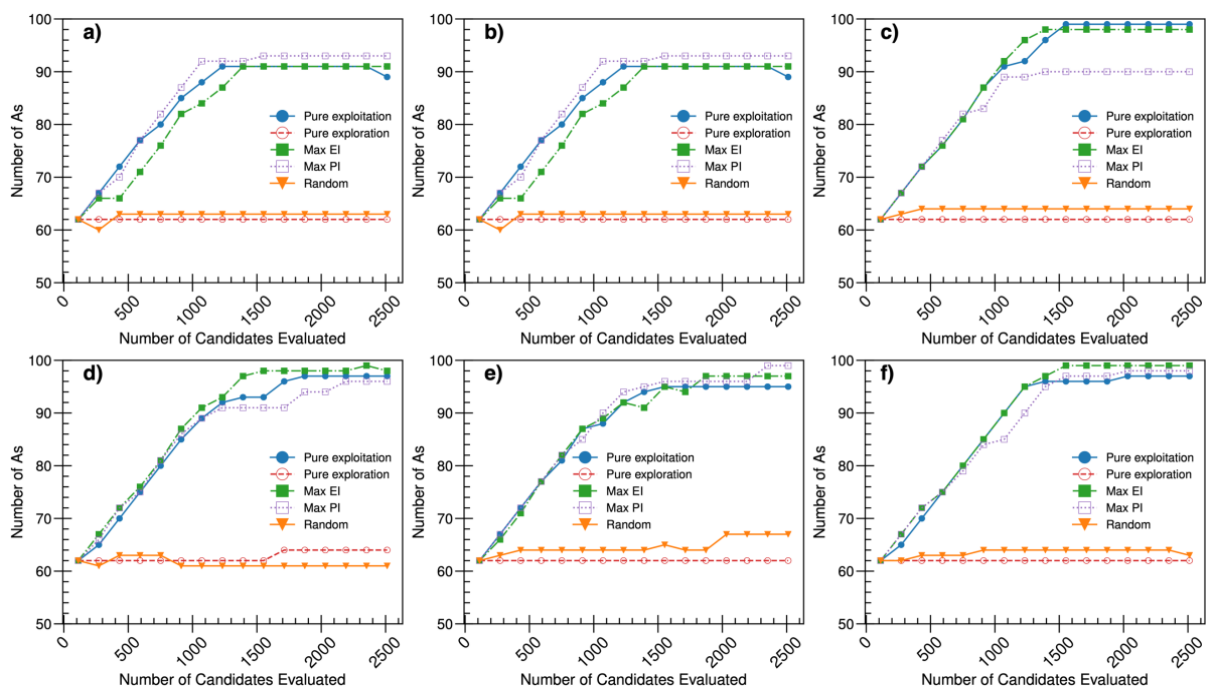

*Figure 4: Quality of solution obtained via active learning. The number of A type moieties present in the best polymer identified during each step is plotted for active learning via GPR-Rbf kernel, GPR-Matérn Kernel, SVR-Linear Kernel, SVR-Rbf kernel, KRR-rbf kernel, and KRR-Laplacian kernel in (a), (b), (c), (d), (e) and (f), respectively. The x-axis represent the cumulative MD simulations conducted at given step of the active learning cycle. For each of these six strategies, five different query strategies are used as mentioned in all the subplots.*

## 3.2 Selection of Query Strategy

The impact of query strategies is analyzed for a representative case of GPR model- Matérn kernel. Figure 5 describes the spread of Rg values corresponding to the candidates that are sampled and appended to the training set at each iteration. The initial training set contains sequences whose Rg values span from 2.848 to 5.030. This wide spread in Rg values is reduced to a narrower breadth (spanning across 2.7 to 3.0), at the very first iteration of active learning with the query strategies - Pure Exploitation, Max EI and Max PI. This exemplifies the brilliance of these query strategies in selectively eliminating the sampling of sequences with higher Rg values. The newer sequences that are sampled in subsequent iterations exhibit a fairly constant/mildly diminishing breadth of the spread in Rg values, eventually resulting in a very narrow spread at the final iteration. For the pure exploration query strategy and the random selection, the Rg values span across 2.5 to 5.5 for the entire range of iterations. We note that the estimation of predictions and uncertainties (when the initial training set is fixed) by GPR do not have any dependency on the randomized bootstrap sampling, unlike SVR or KRR models. In SVR and KRR, for a particular query strategy, if the same initial training set is retained and the optimizer run multiple times, one might observe slightly different results every time owing to the differences in the bootstrap sampling at every iteration. If one is extremely intent to eliminate the influences in optimizer performance arising out of variations in bootstrap sampling, GPR would be the best approach.

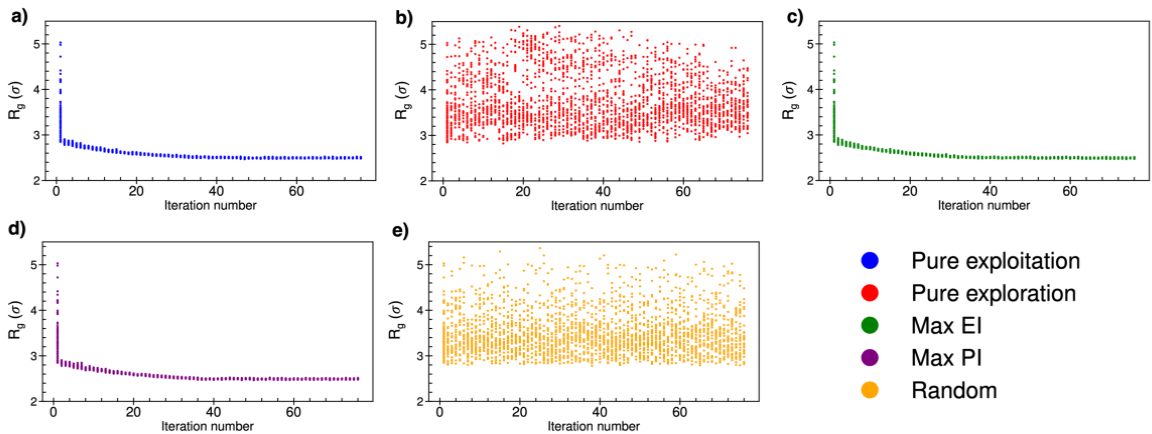

*Figure 5: Convergency of query strategies. The variation of Rg of the 32 candidates that are picked by a query strategy during an active learning cycle is shown as a function of the iteration number, for a specific surrogate model (Gaussian Process Regression with Matérn kernel) with query strategies, Pure exploitation, Pure exploration, Max EI, Max PI, Random in (a), (b), (c), (d), and (e), respectively*

## 3.3 Stopping Criteria for Active Learning

The stopping criteria, i.e., the precise condition/circumstance at which the iterative process of active learning would be ceased is an important aspect of any optimization algorithm. The general notion is to stop subsequent sampling when one is satisfied with the outcome of the target property or when one has run out of computational budget[16]. In this work, for all the sub-cases, the number of iterations was taken to be 75, and for most of them, the desired Radius of Gyration was obtained within 75 iterations. For the first problem statement, owing to our theoretical understanding, we know that the minimum Rg would be found in a sequence that is predominantly populated by As, and hence the desired solution was not unknown. For problems involving longer sequences with increased number of monomers wherein the desired solution is unknown, each computation would be far more expensive, and defining a stopping criterion becomes inevitable in order to save on computational budget, and to avoid excessively scanning the sequence space. In such cases, one could consider a heuristic measure such as saturation in the marginal improvement of the optimizer over a considerable number of iterations (say 10-20 iterations). In Figure 6, the incremental improvement in the performance of the optimizer is quantified for different query strategies by plotting the extent of decrease in Rg as a function of the number of iterations. For the query strategies - Pure Exploitation, Max EI, and Max PI, the extent of decrease in Rg is high in the initial iterations, and becomes marginally small or fairly constant towards higher iterations (after 40-50 iterations, corresponding to roughly 1400-1700 computations). For the Pure Exploration query strategy

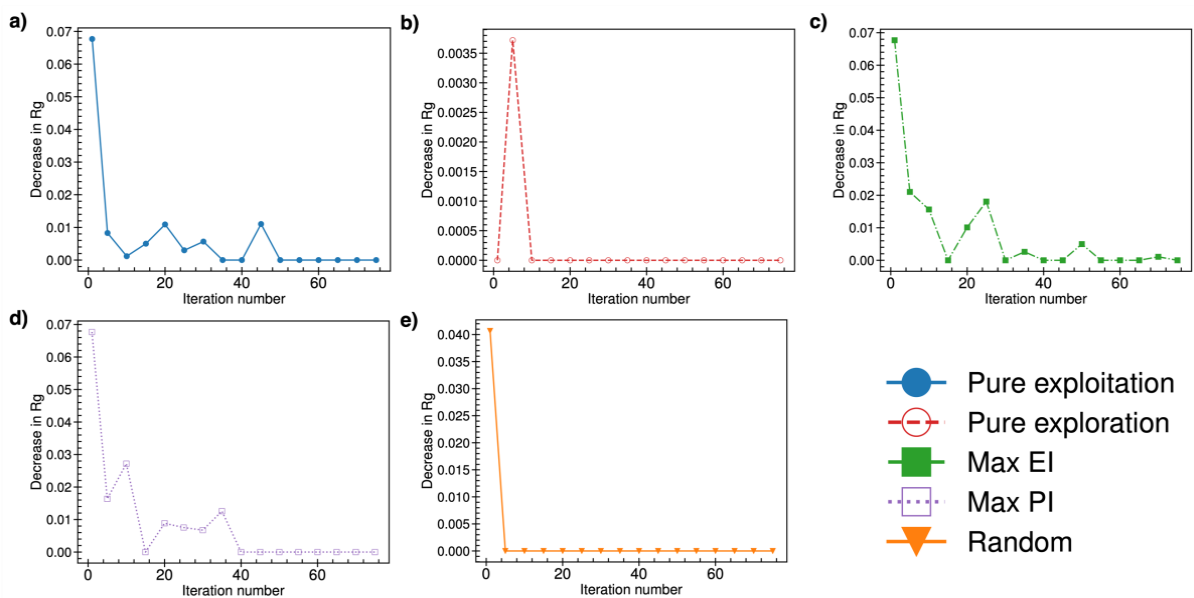

*Figure 6: Analysis of Stopping Criteria for different query strategies. The decrease in Rg is plotted as a function of the iteration number, for a specific surrogate model (Gaussian Process Regression with Matérn kernel)*

and the random selection approach, the extent of decrease in Rg becomes zero after a very few iterations, owing to the optimizer being stuck in a local optimum at a very early stage in the active learning cycle. Thus, it is important to consider the notion of incremental improvement cessation along with the larger perspective of convergence to the desired target property value, so as to avoid pre-mature or incorrect inferences. A similar approach of quantifying the rate of improvement has been commonly used in literature by means of an opportunity cost[9] concept which defines the modulus difference between the overall best value in the data set and the best-so-far. Balachandran et al.[17] have formulated stopping criterion by tracking the max Expected Improvement (EI) at the end of each iteration. The recommendation is to not stop the iterative cycle immediately after max EI has reached a value of zero, but to run the iterative active learning loop for further iterations until max EI becomes consistently zero and does not further increase.

**3.4 Assessment of Surrogate Models' Quality during Active Learning**

$R^2$ in the Figures 7 stands for the coefficient of determination computed with the predicted and true values of Radius of gyration for the 32 newly sampled sequences at the end of every iteration. $R^2$ is computed as $1 - \frac{SS_{res}}{SS_{tot}}$ wherein $SS_{res}$ stands for the residual error and $SS_{tot}$ stands for the total error. $R^2$ measures the goodness of fit of a model with that of a horizontal straight line. When $\frac{SS_{res}}{SS_{tot}}$ is a small positive fraction, the model fitting is very good and this yields an $R^2$ value close to 1. When $SS_{res} > SS_{tot}$, the model fitting is worse than a horizontal that of a horizontal line, and $R^2$ ends up becoming negative. Figure 7 compares the trends in $R^2$ for a specific case of Gaussian Process Regression surrogate model (Matérn kernel) between pure exploitation and pure exploration query strategies.

There is high likelihood to observe high model error and poor $R^2$ in problems with relatively small datasets and a vast search space. For the query strategy - pure Exploitation, $R^2$ is significantly negative across the 75 iterations indicating that the model is fairly inaccurate. Yet, convergence towards optimal sequences is ensured. Interestingly, certain studies[9] have documented similar observations of how the adaptive design is quite forgiving of the quality of the surrogate model (even in the regime of large model error) as long as appropriate query strategies are employed to find optimal solutions. While the importance of model quality in the Active Learning is an aspect that needs a lot of investigation and rigorous analysis, it can be safely stated that the iterative approach of active learning with the pure exploitation to sample newer candidates fairly guarantees the performance of the optimizer in this study. On the other hand, for the query strategy pure exploration which does not efficiently identify sequences $R^2$ is positive and fairly close to 1 across the 75 iterations. This observation is expected, as the very intent of the pure exploration query strategy is to improve the model quality by continuously appending the training set by sampling data points which are predicted to have high uncertainty.

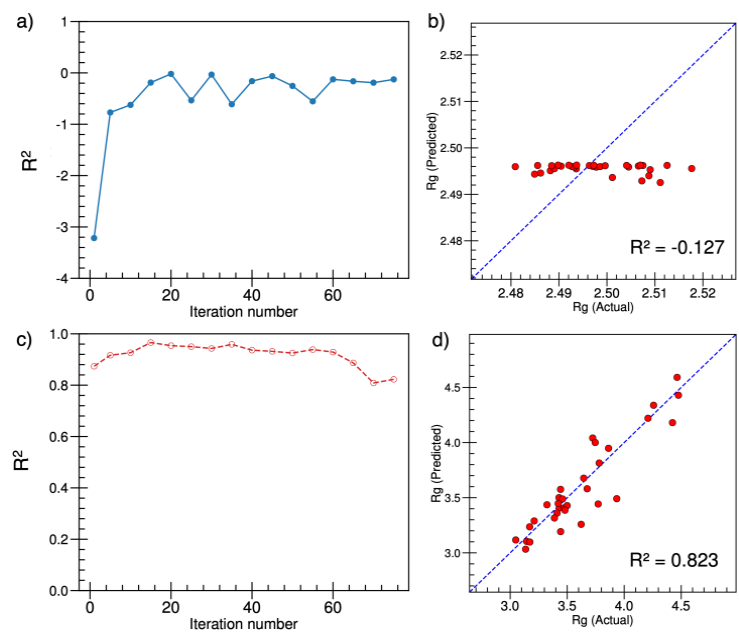

*Figure 7: Plot of the R-squared (comparing true and predicted values of the 32 candidates appended to the training set in each iteration) as a function of iteration number for active learning via (a) pure exploitation (c) pure exploration. Plot of predicted values versus true values at the 75th iteration for active learning via (b) pure exploitation (d) pure exploration*

### 3.5 Trade-off between Exploration and Exploitation

While the premise of using a query strategy such as Max EI or Max PI is to balance the trade-off between exploitation and exploration, it can be inferred from the graphs of Figure 3 that the trends in Rg values sampled by these query strategies is quite similar to that of pure exploitation, and very dissimilar to pure exploration. In order to delve deeper to understand this observation, the following procedure was attempted. At every iteration of sampling 32 Rg values by the Max EI query strategy, in parallel, the list of sequences that are recommended (based on the existing training set, until that particular iteration) by a purely exploitative and a purely explorative approach were recorded. The number of candidate sequences that are recommended in common by the max EI and pure exploration, and max EI and pure exploration are assessed as a function of number of iterations. The same set of steps are done in the case of the query strategy Max PI as well. The rationale behind adopting this procedure is to formulate an apples-to-apples comparison. Comparing the common sequences recommended by the two query strategies would be an ill-defined comparison on account of the differing cumulative training sets at the end of every iteration. From Figures 8a and 8c, it can be observed that the number of candidate sequences recommended in common by pure exploitation and max EI, and pure exploitation and max PI are quite high, to the extent of 15-25 number of sequences (out of 32) being recommended similarly, in the initial 30-40 iterations. On the other hand, it can be inferred from Figures 8b and 8d, that there are zero sequences recommended in common by the pure exploration query strategy vis-à-vis Max EI or Max PI. Considering the vast sample space of $2^{99}$ sequences, this observation cannot be a mere coincidence, and thus the explanation is that μ dominates over σ[17] in the expression $z = \frac{\mu(x_{test}) - (\mu^* + \epsilon)}{\sigma(x_{test})}$, leading to a predominant exploitation-based approach for a large number of initial iterations. The number of sequences recommended in common with the pure exploitation approach by these two query strategies becomes smaller in higher iterations, and though there is not so much of a commonality with sequences recommended by pure exploration, this could possibly explain the mildly exploratory behaviour that operates at higher iterations, facilitating the escape from local optima. This is counter-intuitive to the general notion of exploitation-exploration trade-off which attempts to explore the search space predominantly in the initial iterations, and subsequently works on exploitation to converge towards the desired target property. This analysis suggests that a pure exploitation or a combination of exploitation and exploration lead to convergence with equal efficiency and produce high quality solutions. Evidentially, a design protocol with only exploration produces poor quality solutions.

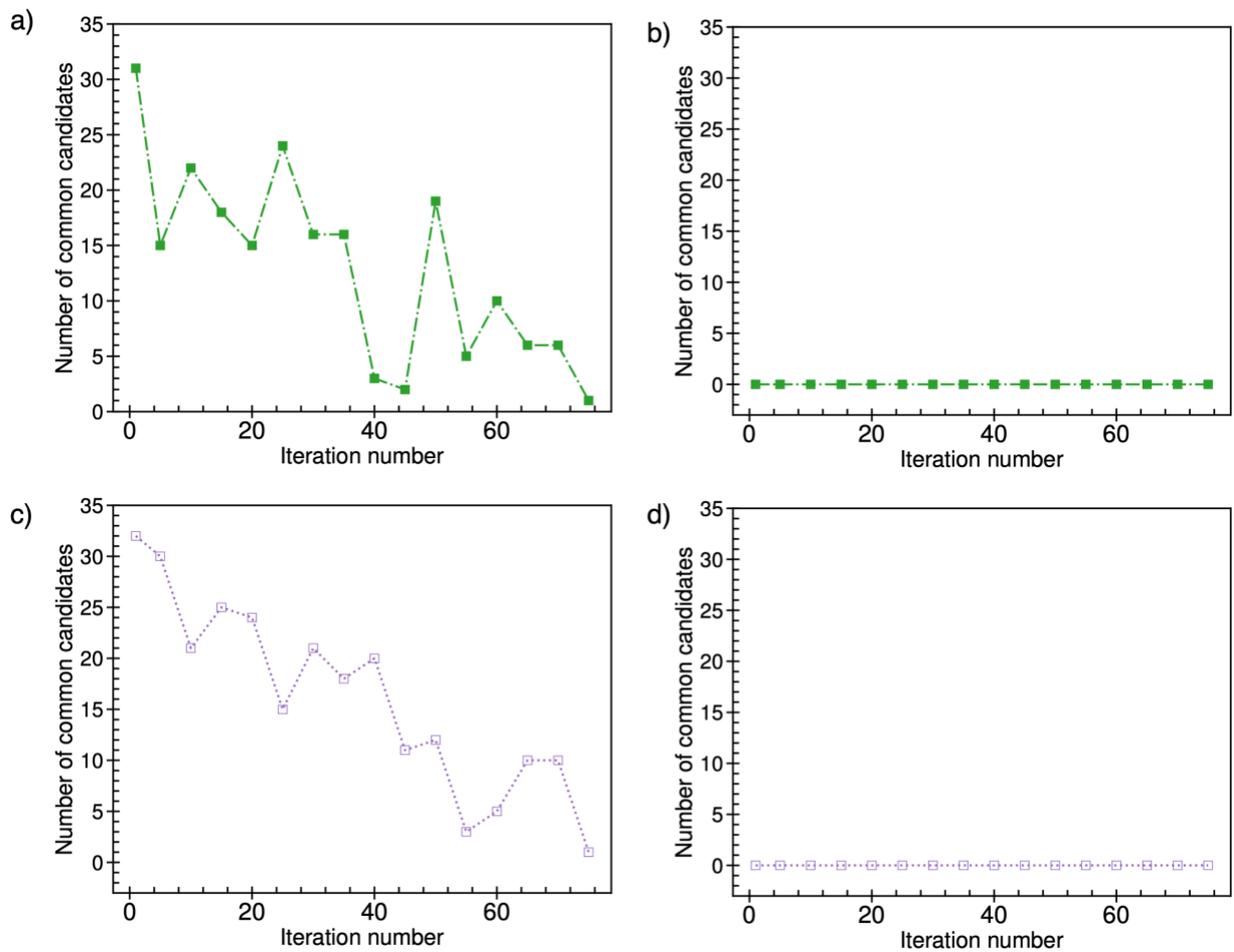

*Figure 8: Analysis of Exploitation-Exploration Trade-off. Number of common candidates among Max EI – pure exploitation, max EI-pure exploration, max PI – pure exploitation, max PI- pure exploration are shown in (a), (b), (c) and (d), respectively.*

## 3.6 Selection of New Candidates during Active Learning

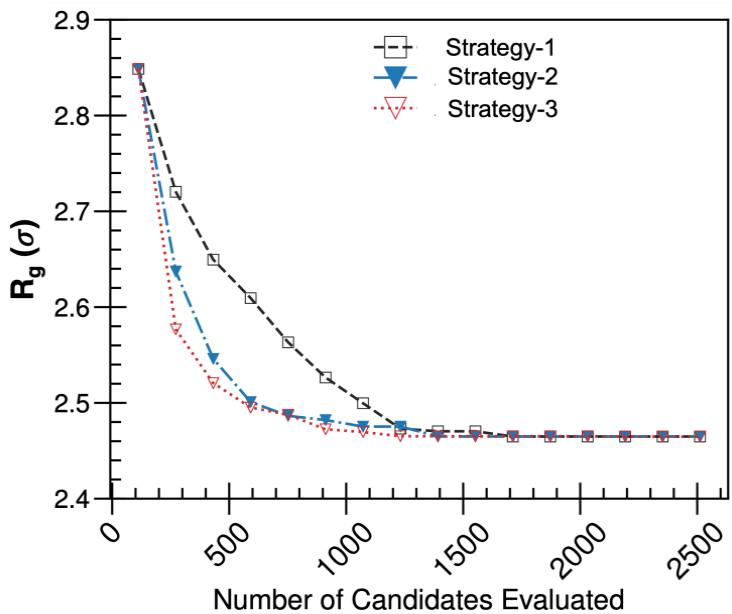

*Figure 9: Performance of active learning for different strategies of selecting new candidates. The radius of gyration of the best candidate in each iterations is plotted as a function of total number of candidates cumulatively screened via MD simulations during an active learning cycles.*

At each iteration, a small search space is defined by enumerating the possible sequences that are sequentially proximate to the sequences in the training set accumulated until that particular iteration. For every copolymer sequence in the training set, the exhaustive list of copolymer sequences that differ in the monomeric entity at exactly one position (i.e., one position of A replaced by B, or one position of B replaced by A) is considered to populate the search space. Here, we refer this strategy as strategy-1. While such a candidate selection strategy indeed facilitates the convergence towards optimal candidates as show in Figure 3, the convergence can be accelerated by defining various selection methods. For example, we test two more strategies for candidate selection. We define strategy-2 where randomly selected four moieties change their types, and in strategy-3 involve flipping the type of 10 moieties in the chain. We repeat the active learning cycle for these two new strategies and compare their performance with that of strategy-1 for a specific case of GPR surrogate model with Matérn kernel and Max PI query strategy. As shown in Figure 9, the performance of strategy-1 and strategy-2 are very identical. However, the active learning processing is significantly accelerated in strategy-2 and strategy-3 in comparison with strategy-1.

**3.7. Protein-like Sequence Design**

Based on the inferences drawn from implementing Bayesian Optimization for a copolymer sequence design problem with no constraints on the number of As and Bs, Gaussian Process Regression (Matérn kernel) and Max. PI (Probability of Improvement) were chosen as the surrogate model and query strategy respectively for the subsequent problem - identification of optimal copolymeric sequences with least radius of gyration value with As and Bs constrained to be in a 50:50 ratio. Similar to the previous problem, the optimal solution is identified after initializing the optimizer with a randomly chosen training set and running the optimizer for 75 iterations. Unlike the first problem in which the solution is intuitively known, the optimal sequences obtained in this problem have a non-intuitive, non-periodic configuration. While, it would be hard to comment on the characteristics of a polymer sequence by merely reading the sequence configuration, the polymer conformations throw light on the spatial arrangement of monomers A and B and the associated connection with a substantially low Radius of Gyration value. In addition, radial probability plots were also constructed in order to better understand the spatial arrangements. As an illustrative example, the radial probability plot and the conformational structure of a randomly chosen sequence in the initial training set are compared with that of the optimal sequence identified after 75 iterations in Figure 10. While the initial sequence (with Rg = 3.679 $\sigma$) has no clear distinct monomer-A dominated-zones or monomer-B dominated zones, the optimal sequence (with Rg = 2.856 $\sigma$) demonstrates a distinct difference in the dominance of monomer units A and B at different radial distances. In the optimal sequence, though there are regions where both monomers A and B are likely to co-exist, it can be clearly seen that monomer A is concentrated in the core, while monomer B is in the periphery. The conformational structure of the optimal sequence has a dense core dominated by monomer A with the periphery dominated by monomer B, resulting in a spherical globule-like structure. Such a distinct core-shell framework with clear-cut domination by different chemical species is quite resembling of secondary structures of certain proteins, which have a packing of hydrophobic moieties in the core and polar groups on the periphery. Thus, by performing merely 2512 MD computations, the Active Learning framework efficiently scans the astronomically large search space and identifies the optimal sequence with a distinct core-shell conformation.

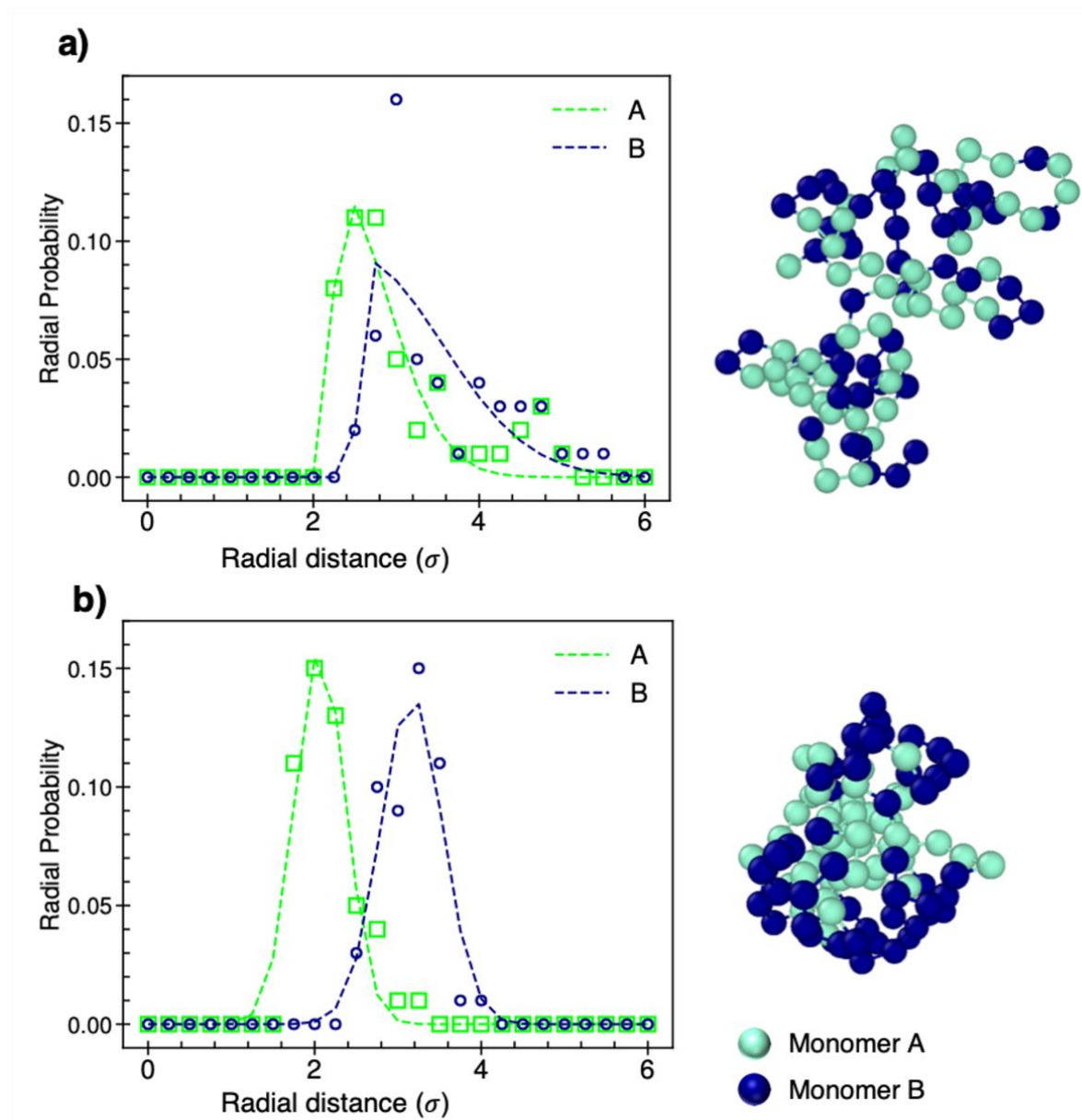

*Figure 10: Protein-like aggregate. Density profile of A and B type moieties from the centroid of the globule for a randomly chosen sequence and active learning identified sequence for a chain length of 100 beads with equal proportion of A and B is shown in (a) and (b), respectively. The respective MD snapshots are shown in the right side of the figures.*

## 5. Conclusions

Active learning has emerged as an attractive tool for materials design. The performance of an active learning framework depends on large number of factors including choices of surrogate model, initial data set, query strategy, uncertainty quantification, selection of search space, stopping criteria. The implications would be profound for problems with increasing computational complexity, wherein one could reap enormous savings in computational budget by avoiding excessive exploration of the feature space of a material. Here, we survey some of these critical aspects of active learning for a sequence design problem. We consider a model copolymer with two chemically dissimilar moieties and conduct coarse-grained molecular dynamics simulation to compute its radius of gyration in an implicit solvent condition. This CGMD simulation is integrated within an active learning method for inverse design of copolymer sequence with target radius of gyration. We first identify optimal strategies and parameters by targeting a known solution. Subsequently, we use the best active learning strategy for an unknown problem wherein we identify copolymer sequence that produce core-shell structure similar to protein folding.

It has been observed that the active learning framework is forgiving of the quality of the surrogate model, as long as effective query strategies are employed. The systematic search space expansion strategy with its foundation on conditional probability ensures the robustness of the convergence to optimality irrespective of the variations in the initial training set. However, the appropriate choice of the combination of surrogate model and query strategy is very important for high fidelity solution in any material design problem that attempts to use active learning. The best combination also depends on the size of the training data, the complexity of the search space, and a host of other factors, and, indeed, there is no universal choice of optimizer for an active learning framework. The best that can be stated is that the performance of the Bayesian Optimizer can be guaranteed to be superior to random sampling of the search space for most problems. Further attempts could potentially be focused on utilizing different surrogate model–query strategy combinations at different stages of the iterative construct. Hyper-parameter selection is another domain which needs to be investigated in greater details as it can potentially fine-tune the performance of the optimization algorithms. The estimates of hyper-parameter selection whether it is by means of optimizing log marginal likelihood (in the case of GPR) or by Bayesian optimization in the hyper-parameter space (in the case of SVR and KRR) could potentially be fine-tuned by considering a larger initial training set, or considering multiple training sets. The technique of bootstrap sampling, method of search space definition, the notion of stopping criterion and a few other aspects could all be formulated differently. The bootstrap sampling with replacement approach that has been implemented in this work with the number of bootstrap samples as 10, could be extended to a bootstrap sample size of 100 or 1000, as use of an ensemble with greater number of bootstrapped samples is known to mimic the normal distribution to a greater extent. Defining a search space at each iteration could be done differently, by working with notions other than swapping, and such approaches can potentially accelerate the convergence of the optimizer. With increasing advancements in the interface of the fields of data analytics, information sciences and material sciences, it can be expected that advanced algorithms would enhance the robustness of the optimizer, achieve more effective circumvention of suboptimal traps in a material's configurational space, facilitate accelerated convergence to optimal solutions and ultimately result in tremendous savings on computational costs.

**Acknowledgement**

The work is made possible by financial support from SERB, DST, Gov of India through a start-up research grant (SRG/2020/001045) and National Supercomputing Mission's research grant (DST/NSM/R&D_HPC_Applications/2021/40). This research used resources of the Argonne Leadership Computing Facility, which is a DOE Office of Science User Facility supported under Contract DE-AC02-06CH11357. We also used computational facility of the Center for Nanoscience Materials. Use of the Center for Nanoscale Materials, an Office of Science user facility, was supported by the U.S. Department of Energy, Office of Science, Office of Basic Energy Sciences, under Contract No. DE-AC02-06CH11357.We acknowledge the use of the computing resources at HPCE, IIT Madras.